%% file: MiNNLO_bbZ.tex
\documentclass[longbibliography,nofootinbib,twocolumn,preprintnumbers]{revtex4-1}

\pdfoutput=1
\usepackage{amsmath}
\usepackage{amssymb}
\usepackage{graphicx}
\usepackage{booktabs}
\usepackage{color}
\usepackage{bbold}
\usepackage{amsmath}
\usepackage{tabularx}
\usepackage{multirow}
\usepackage{physics}
\usepackage[colorlinks=true
  ,urlcolor=blue
  ,anchorcolor=blue
  ,citecolor=blue
  ,filecolor=blue
  ,linkcolor=blue
  ,menucolor=blue
  ,pagecolor=blue
]{hyperref}
\usepackage{tikz-feynman}
\newcommand{\mathd}{\mathrm{d}}

\newcommand{\as}{\alpha_s}

\newcommand{\ptvec}{{\vec{p}_T}}
\newcommand{\pt}{p_T}

\newcommand{\noun}[1]{{\scshape #1}}

\newcommand{\POWHEG}{\noun{POWHEG}}

\newcommand{\POWHEGBOXRES}{\noun{POWHEG-BOX-RES}}

\newcommand{\minlo}{{\noun{MiNLO$^{\prime}$}}}
\newcommand{\minnlo}{{\noun{MiNNLO$_{\rm PS}$}}}

\newcommand{\PYTHIA}[1]{\noun{Pythia{#1}}}

\newcommand{\OpenLoops}{{\noun{OpenLoops}}}

\newcommand{\citere}[1]{Ref.\,\cite{#1}}
\newcommand{\citeres}[1]{Refs.\,\cite{#1}}
\newcommand{\eqn}[1]{Eq.\,(\ref{#1})}

\newcommand{\fig}[1]{Figure\,\ref{#1}}

\newcommand{\tab}[1]{Table\,\ref{#1}}

\newcommand{\app}[1]{Appendix\,\ref{#1}}

\graphicspath{{figures/}}

\AtBeginDocument{
\heavyrulewidth=.08em
\lightrulewidth=.05em
\cmidrulewidth=.03em
\belowrulesep=.65ex
\belowbottomsep=0pt
\aboverulesep=.4ex
\abovetopsep=0pt
\cmidrulesep=\doublerulesep
\cmidrulekern=.5em
\defaultaddspace=.5em
}
\begin{document}

\title{Next-to-next-to-leading order event generation for \\
       Z-boson production in association with a bottom-quark pair}

\preprint{MPP-2024-54 --- PSI-PR-24-10 --- ZU-TH 15/24}

\author{Javier Mazzitelli$^1$, Vasily Sotnikov$^2$, Marius Wiesemann$^3$\vspace{1em}}

\affiliation{$^1$ Paul Scherrer Institut, CH-5232 Villigen PSI, Switzerland}
\affiliation{$^2$ Physik Institut, Universit\"at Z\"urich, CH-8057 Z\"urich, Switzerland}
\affiliation{$^3$ Max-Planck-Institut f\"ur Physik, Boltzmannstraße 8, 85748 Garching, Germany}

\begin{abstract}
\noindent
We consider the production of a $Z$ boson decaying to leptons in association with a bottom-quark pair 
in hadronic collisions. For the first time, we compute predictions at next-to-next-to-leading order (NNLO)
in QCD, and we combine them with the all-orders radiative corrections from a parton-shower simulation (NNLO+PS).
Our method represents the first approach to NNLO+PS event generation applicable to processes featuring a colour singlet
and a heavy-quark pair in the final state. The novel two-loop corrections are computed for massless bottom quarks,
and the leading mass corrections are restored through a small-mass expansion. The calculation is carried out
in the four-flavour scheme, and we find that the sizeable NNLO QCD corrections lift the long-standing tension between
lower-order predictions in four- and five-flavour schemes. Our predictions are compared to a CMS measurement for $Z$ boson
plus $b$-jet production, achieving an excellent description of the data.
\end{abstract}

\maketitle

{\it Introduction.---}%
The quest to uncover signals of new physics at the Large Hadron Collider (LHC)
has become a major endeavour within the extensive physics program of the LHC experiments,
encompassing both the direct detection of resonances and the analysis of small deviations
from the Standard Model (SM) through precision measurements. 
The success of the latter critically depends 
on the availability of highly accurate simulations and predictions for all classes of LHC reactions.

The associated production of a $Z$ boson and a bottom-quark pair  ($b\bar{b}Z$) 
plays a special role in the LHC physics program.
For one, it yields heavy-quark mass effects to Drell-Yan production, a process that is measured very 
accurately at hadron colliders and 
used to extract the mass of the $W$ boson, see e.g.\ \citeres{ATLAS:2017rzl,LHCb:2021bjt,CDF:2022hxs}.
Moreover, $b\bar{b}Z$ production is a major background to $ZH$ 
measurements~\cite{ATLAS:2020fcp,CMS:2023vzh} and to various BSM searches, see e.g.\ \citere{CMS:2023byi}.

Precision measurements of $Z$-boson production
with bottom-flavoured jets ($b$-jets) have been performed both at the 
Tevatron~\cite{CDF:2008vda,D0:2010acs,D0:2013cga} and by the LHC experiments~\cite{CMS:2014jqj,ATLAS:2014rjv,LHCb:2014ydc,CMS:2016gmz,ATLAS:2020juj,CMS:2021pcj,ATLAS:2022uav,ATLAS:2024tnr}.
So far, there has been a significant discrepancy among the $Z$+$b$-jet theory predictions, depending on
whether the bottom quark is assumed to be massive in a four-flavour scheme (4FS) 
or massless in a five-flavour scheme (5FS) calculation~\cite{Krauss:2016orf}. 
Especially, state-of-the-art 4FS predictions are in
tension with measurements of $Z$+$b$-jet production~\cite{CMS:2014jqj,ATLAS:2014rjv,CMS:2016gmz,Krauss:2016orf,ATLAS:2020juj}.

Next-to-leading-order (NLO) corrections in QCD have been computed 
both in the 4FS~\cite{FebresCordero:2008ci} (see also \citere{Campbell:2000bg}) and in the 5FS \cite{Campbell:2003dd}.
In the 5FS, $Z$ boson production with a single $b$-jet is known through next-to-next-to-leading-order (NNLO) \cite{Gauld:2020deh}.
\citeres{FebresCordero:2008ci,FebresCordero:2009xzo} showed that the inclusion of bottom-mass effects in the 4FS
is important for the $b\bar{b}Z$ cross section and in kinematic distributions of the bottom quarks for scales close to $m_b$.
The matching of the NLO corrections with a parton shower (NLO+PS) is available in both schemes 
within {\sc MadGraph5\_aMC@NLO} \cite{Frederix:2011qg} and within {\sc Sherpa} \cite{Krauss:2016orf}, where the 
5FS implementation in {\sc Sherpa} includes up to two extra $b$-jets at NLO+PS via multi-jet merging.
The combination of four- and five-flavour predictions at NLO+PS was achieved in \citere{Hoche:2019ncc},
see also \citere{Forte:2018ovl} for an earlier study. 
Higher-order corrections to $b\bar{b}Z$ production are substantial, especially in the 4FS, with very sizeable perturbative uncertainties.
Prior to this work, no NNLO results had been obtained for any of the flavour schemes.

In this letter we considerably improve the accuracy of $Z$+$b$-jet theory predictions 
by computing the NNLO QCD corrections to $b\bar{b}Z$ production for massive bottom quarks.
Our results are implemented in a fully exclusive Monte Carlo 
event generator by matching the NNLO corrections to a parton shower (NNLO+PS).
Our approach represents the first NNLO+PS method for such an involved final state, featuring 
both a heavy-quark pair and a set of colour-singlet particles.
The two-loop corrections for the $b\bar{b}Z$ process are presented here for the first time,
by exploiting the massless amplitudes~\cite{Abreu:2021asb} to approximate the massive ones.
We use our novel results to investigate whether the NNLO corrections resolve the long-standing discrepancies
between 4FS and 5FS predictions, 
and compare our results with the latest $Z$+$b$-jet measurement by CMS~\cite{CMS:2021pcj}
to asses their impact on the tension of previous 4FS predictions with LHC data.

\vspace{0.5cm}{\it Methodology.---}%
We consider the process
\begin{align}
pp \to b\bar{b} \ell^+ \ell^- +X\,,
\end{align}
which we denote as $b\bar{b} Z$ production for brevity. Sample Feynman diagrams at the leading order (LO) are shown in \fig{fig:diagrams}.
They involve both quark- and gluon-initiated channels, where the $Z$ boson can be radiated off any of the quark lines.

\begin{figure}[t]
\input{feynman_diagrams.tex}
\end{figure}

Among the NNLO+PS frameworks for colour-singlet
production~\cite{Hamilton:2012rf,Alioli:2013hqa,Hoeche:2014aia,Monni:2019whf,Monni:2020nks}, only the \minnlo{} approach \cite{Monni:2019whf,Monni:2020nks}
has been extended to processes featuring colour charges in initial and final state, namely heavy-quark pair production \cite{Mazzitelli:2020jio,Mazzitelli:2021mmm,Mazzitelli:2023znt}.
We further extend the \minnlo{} method to construct a fully exclusive NNLO+PS generator 
for the associated production of a heavy quark pair ($Q\bar Q$) with a system colour-singlet particles ($F$).

The general structure of large logarithmic contributions at small transverse momentum 
for $Q\bar Q$ and $Q\bar QF$ production is very similar \cite{Catani:2021cbl}.
As a result, we can exploit the analytic formula for the cross section differential in the 
transverse momentum $\pt\equiv|\ptvec |$ and the Born phase space 
$\Phi_{\rm B}$ of the final-state system, 
which was derived in \citeres{Mazzitelli:2020jio,Mazzitelli:2021mmm}
from the well-known factorization theorem for $Q\bar Q$ production at small $\pt$ \cite{Zhu:2012ts,Li:2013mia,Catani:2014qha,Catani:2018mei} and reads
\begin{align}
\label{eq:start}
    \frac{\mathd\sigma}{\mathd{} \pt\mathd{}\Phi_{\rm B}}  
    &= \frac{\mathd{}}{\mathd{}
  \pt}\Bigg\{\sum_{c}\Bigg[\sum_{i=1}^{n_c}{\cal C}^{[\gamma_i]}_{c\bar{c}}e^{-\tilde{S}^{[\gamma_i]}_{c\bar{c}}}\Bigg]{\cal L}_{c\bar{c}} \Bigg\}+R_f\hspace{0.01cm}.
\end{align}
The first term includes the singular (and constant) contributions in $\pt$,
which represents the transverse momentum of the $Q\bar{Q}F$ system here,
while $R_f$ adds the regular contributions.

This formula retains NNLO plus LL accuracy, and it preserves the class of NLL corrections  
that  traditional PS algorithms~\cite{Catani:1990rr} include.
Therefore, it is the starting point to build a Monte Carlo
algorithm for the generation of NNLO events within the \POWHEG{} framework
\cite{Alioli:2010xd,Jezo:2015aia}, which has been demonstrated several 
times \cite{Monni:2019whf,Monni:2020nks,Mazzitelli:2021mmm}, most recently for $Q\bar{Q}$ production in \citere{Mazzitelli:2021mmm}.
In the following, instead, we focus on the main changes from $Q\bar Q$ to $Q\bar QF$ processes.

The sum over $c$  in \eqn{eq:start} runs over the flavour
configurations $\{g,q,\bar q\}$ of the incoming partons of flavour $c$ and $\bar c$,
while the sum over $i$ originates from diagonalizing the one-loop soft anomalous 
dimension $\mathbf{\Gamma}^{(1)}$
with complex coefficients ${\cal C}^{[\gamma_i]}_{c\bar{c}}$,
where $n_c$ is determined by the SU(3) representation of a given flavour configuration.%
\footnote{The notation $X^{(i)}$ denotes the $i$-th term in the perturbative expansion of the quantity $X$ with respect to $\alpha_s/(2\pi)$.}
The eigenvalues of $\mathbf{\Gamma}^{(1)}$, denoted by $\gamma_i$, induce a modification
to the Sudakov exponent, as indicated by the notation $\tilde{S}_{c\bar c}^{[\gamma_i]}$.
We note that $\mathbf{\Gamma}^{(1)}$, its eigenvalues, and the coefficients stemming from its diagonalization have the same form as for $Q\bar{Q}$ production (see Appendix\,A of \citere{Mazzitelli:2021mmm})
after adapting the process-dependent LO matrix 
$\mathbf{H}^{c\bar{c}}$, which enters the definition of ${\cal C}^{[\gamma_i]}_{c\bar{c}}$. We construct this matrix from colour-decomposed scattering amplitudes of \OpenLoops{}~\cite{Cascioli:2011va,Buccioni:2017yxi,Buccioni:2019sur}.
Contributions from ${\mathbf \Gamma}^{(2)}$ are included 
in the Sudakov with an approximation of its corrections beyond NNLO.
The general form of ${\mathbf \Gamma}^{(2)}$
includes terms proportional to three parton correlations,
whose expectation value with the LO matrix element vanishes in the $Q\bar{Q}$ case. 
For $Q\bar{Q}F$ processes they contribute, and we therefore include them in our implementation.

The last ingredient of \eqn{eq:start} is the luminosity factor 
\begin{equation}
\label{eq:lumi}
{\cal L}_{c\bar{c}} \equiv \frac{|M^{(0)}_{c\bar{c}}|^2}{2 m_{\scriptscriptstyle Q\bar{Q}F}^2}\sum_{i,j}\left[{\rm Tr}({\tilde{\mathbf H}}_{c\bar{c}}{\mathbf D})(\tilde{C}_{ci}\otimes f_i)
  \,(\tilde{C}_{\bar{c} j}\otimes f_j)\right]_\phi \hspace{-0.1cm}
\end{equation}
with 
the LO matrix element $M^{(0)}_{c\bar{c}}$, 
the invariant mass of the $Q\bar QF$ system $m_{\scriptscriptstyle Q\bar QF}$,
and the convolution, denoted by the operation $\otimes$, of the collinear coefficient functions
$\tilde C_{ij}$ with the parton densities $f_i$.
Quantities in bold face are operators in colour space, 
and the trace ${\rm Tr}({\tilde{\mathbf H}}_{c\bar c}{\mathbf D} )$ runs over the colour indices.
The symbol $[\cdots]_\phi$ denotes the average over the azimuthal angle $\phi$ of
$\ptvec$, and the operator of azimuthal correlations ${\mathbf D}$
is defined such that $[{\mathbf D}]_\phi={\mathbb 1}$.

The notation
${\rm Tr}({\tilde{\mathbf H}}_{c\bar{c}}{\mathbf D})({\tilde C}_{ci}\otimes f_i)
\,({\tilde C}_{\bar{c} j}\otimes f_j)$ is symbolic. In particular, the gluon-initiated channel features
a richer Lorentz structure, typically denoted by additional
$G$ functions~\cite{Catani:2010pd}, which encode
azimuthal correlations of collinear origin and are kept implicit here.
Therefore, the product of
${\rm Tr}({\tilde{\mathbf H}}_{c\bar{c}}{\mathbf D})$ and
$({\tilde C}_{ci}\otimes f_i) \,({\tilde C}_{\bar{c} j}\otimes f_j)$
should be understood as a tensor contraction, which leads
to additional azimuthal correlations, see \citere{Catani:2014qha}. 
These contributions enter our calculation as terms proportional to
$\langle M_{gg}^{(0)}|\mathbf{D}^{(1)}|M_{gg}^{(0)}\rangle\times G^{(1)}$.
They were computed analytically for $Q\bar Q$ production in \citere{hayk}.
For $Q\bar Q F$ processes we compute them through
a numerical integration over the azimuthal angle in conjugate space ($b$ space).
Similarly, we adapt the implementation of the $G^{(1)}\times G^{(1)}$ 
contribution, which we extract using \OpenLoops{}.

After accounting for the azimuthal correlations in the trace with the
hard-virtual operator $\tilde{\mathbf H}_{c\bar{c}}$ in \eqn{eq:lumi}, we 
 evaluate its contribution as
\begin{align} \label{eq:hard_function}
H_{c\bar c} =
\frac{\ev{\bar{\mathbf{h}}^\dagger \bar{\mathbf{h}}}{\mathcal{R}_{c\bar{c}}}}{\braket*{\mathcal{R}^{(0)}_{c\bar{c}}}}, \quad \bar{\mathbf{h}}\ket{\mathcal{R}_{c\bar{c}}} = \bar{\mathbf{h}}\,\mathbf{Z}^{-1} \ket{\mathcal{M}_{c\bar{c}}},
\end{align}
where $\mathcal{R}_{c\bar{c}}$ is the finite remainder obtained from the virtual amplitude $\mathcal{M}_{c\bar{c}}$ after removing the IR divergences 
through the operator $\mathbf{Z}^{-1}$ \cite{Becher:2009cu,Becher:2009qa}, $\braket*{\mathcal{R}^{(0)}_{c\bar{c}}} = |M^{(0)}_{c\bar{c}}|^2$,
and the operator $\bar{\mathbf{h}}$ determines parton emissions of soft origin, which are known for 
$Q\bar{Q}$ production up to NNLO \cite{Catani:2023tby}. Here, we employ their extension to the general kinematics of $Q\bar Q F$ production \cite{shark:inprep}.

The process-dependent pieces entering the hard function are the finite remainders up to two loops. 
We obtain the one-loop amplitude entering $H_{c\bar{c}}^{(1)}$ and one-loop squared amplitude entering $H_{c\bar{c}}^{(2)}$ from \OpenLoops{}. 
While all other contributions are taken exactly, we calculate
the two-loop finite remainder $2\Re\braket*{\mathcal{R}^{(0)}_{c\bar c}}{\mathcal{R}^{(2)}_{c\bar c}}$ entering $H_{c\bar{c}}^{(2)}$
in an approximation, since its full computation is well beyond current technology for two-loop 5-point amplitudes.
Realizing that there is a hierarchy in the bottom-quark mass $m_b$, we can perform 
an expansion around small $m_b$ of the two-loop amplitude, capturing the constant and logarithmically enhanced terms in $m_b$, while omitting power corrections in $m_b$:
\begin{align}
\label{eq:mbexpansion}
  \frac{2 \Re\braket*{\mathcal{R}^{(0)}_{c\bar{c}}}{\mathcal{R}^{(2)}_{c\bar{c}}}}{\braket*{\mathcal{R}^{(0)}_{c\bar{c}}}} &= \frac{2 \Re\braket*{\mathcal{R}^{(0)}_{0,{c\bar{c}}}}{\mathcal{R}^{(2)}_{0,{c\bar{c}}}}}{\braket*{\mathcal{R}^{(0)}_{0,{c\bar{c}}}}} + \\ \nonumber
  \sum_{i=0}^{4} & \kappa_{c\bar{c},i} ~ \log\qty(\frac{m_b}{\mu_R})^i  ~+~ \order{\frac{m_b}{\mu_h}}, 
\end{align}
where $\mathcal{R}^{(i)}_{0,c\bar{c}}$ denotes the finite remainder of the \emph{massless} $b\bar{b}Z$ amplitude, i.e.\ setting $m_b=0$,
$\mu_R$ is the renormalization scale, and $\mu_h$ is a characteristic hard scale of the process. The process-dependent coefficients $\kappa_{c\bar c,i}$ 
are determined in \app{sec:finite_massification}. They are obtained through a {\it massification} procedure that 
relates the $1/\epsilon^i$ poles of collinear origin in the 5FS with logarithmic terms in $m_b$ in the 4FS \cite{Mitov:2006xs,Wang:2023qbf},
see also \citere{Buonocore:2022pqq} for a recent application to $b\bar{b}W$ production.

In the massless case, the calculation of the two-loop amplitude is still very challenging,
but feasible~\cite{Abreu:2021asb}. 
While the logarithmic terms are reproduced without any approximations, $\Re\langle\mathcal{R}^{(0)}_{0,c\bar{c}}|\mathcal{R}^{(2)}_{0,c\bar{c}}\rangle$ is computed in the 
leading-colour approximation (LCA), with the exception of contributions of $Z/\gamma^\star$ bosons coupling to closed fermion loops, which are omitted.
We have tested the latter 
to be negligible already at the one-loop level (see also \citeres{Dicus:1985wx,Dixon:1997th}).
The LCA is typically accurate within 10\% (see e.g.\ \citeres{Badger:2023mgf,Abreu:2023bdp}).
Since the numerical effect of $\braket*{\mathcal{R}^{(0)}_{0,c}}{\mathcal{R}^{(2)}_{0,c}}$ on the \minnlo{} cross section is typically at the few-percent level, we expect these 
approximations to have a negligible impact on our results.
To calculate  $\Re\langle\mathcal{R}^{(0)}_{0,c\bar{c}}|\mathcal{R}^{(2)}_{0,c\bar{c}}\rangle$ we have implemented a numerical code based on the analytic results of \citere{Abreu:2021asb}, 
employing the \textsc{PentagonFunctions++} code \cite{Chicherin:2020oor,Chicherin:2021dyp,Abreu:2023rco} to evaluate the relevant special functions.

We note that our calculation of the logarithmically enhanced terms in \eqn{eq:mbexpansion} has been rendered possible for 
closed massive fermion loops only by the recent results of \citere{Wang:2023qbf}. 
However, the numerical impact of those contributions turns out to be tiny, at the few-permille level of the NNLO cross section, which we have tested both by excluding them from the two-loop finite remainder and by turning top and bottom loops off in the rest of the calculation.

Since our NNLO+PS generator assumes massive bottom quarks,
a mapping from the massive to the massless phase space is required to evaluate the massless 
remainders $\mathcal{R}^{(i)}_{0,c\bar{c}}$. While different mappings induce only power corrections in $m_b/\mu_h$,
it is mandatory that the mapping avoids the collinear singularities of the massless amplitudes, which in the massive phase space
are prevented by the bottom mass.%
\footnote{We thank Chiara Savoini and Massimilano Grazzini for bringing this to our attention.}
We have tested different mappings 
and found their results to agree at the sub-percent level. The details are given in \app{sec:momentum_maps}.

\vspace{0.5cm}{\it Results.---}%
For the phenomenological study of $b\bar{b}Z$ production at NNLO+PS 
we focus on LHC collisions with $13$\,TeV centre-of-mass energy and
consider the leptonic final states with $\ell=e,\mu$. 
The bottom and top-quark on-shell masses are set to $4.92$\,GeV and $173.2$\,GeV, respectively, with four
massless quark flavours. We employ the corresponding NNLO set of the
NNPDF31 \cite{Ball:2017nwa} parton densities with
$\alpha_s(m_Z)=0.118$.
We use the complex-mass scheme~\cite{Denner:1999gp,Denner:2005fg} and 
the electroweak (EW) input parameters are set in the 
$G_{\mu}$ scheme using \cite{ParticleDataGroup:2020ssz}: $G_F =
1.16639 \times 10^{-5}$\,GeV$^{-2}$, $M_W = 80.385$\,GeV,
$\Gamma_{W} = 2.0854$\,GeV,
$m_{Z} = 91.1876$\,GeV,
$\Gamma_{Z} = 2.4952$\,GeV.
Unless specified otherwise, our default choice for the renormalization scale 
of the two powers of the $\alpha_s$ at Born-level is 
$\mu_R^{(0)}= m_{b\bar{b}\ell\ell}$. The scale of extra powers of  $\alpha_s$
in the radiative corrections and the factorization scale are set following the \minnlo{} prescription~\cite{Monni:2019whf,Mazzitelli:2021mmm}.
We employ the definition of the modified logarithm $L$ in \citere{Mazzitelli:2021mmm}, which smoothly 
turns off resummation effects for $\pt{}$ values larger than $Q=m_{b\bar{b}\ell\ell}/2$.
To avoid the Landau singularity at small $\pt$, the scale of the strong coupling and the parton
densities is smoothly frozen around $Q_0=2$ GeV \cite{Monni:2020nks}.
Scale uncertainties are estimated through the usual 7-point scale variations by a factor of two around the central scale.
As a parton shower we employ \PYTHIA{8}~\cite{Sjostrand:2014zea} with the Monash tune \cite{Skands:2014pea}.

For comparison, we implemented a generator 
for $pp\to b\bar{b}\ell^+\ell^-$ production at NLO+PS 
in the 4FS within \POWHEGBOXRES{} \cite{Jezo:2015aia}. 
In this case we use $m_{b\bar{b}\ell\ell}$ for the central scales.
We also evaluate \minlo{} results, which are NLO accurate for
$b\bar{b}\ell^+\ell^-$ plus zero and one jet, by turning 
off the NNLO corrections in the \minnlo{} generator.

\begin{table}[b!]
  \vspace*{0.3ex}
  \begin{center}
\begin{tabular}{l|cc}
\toprule
& $\sigma_{\rm total}$ [pb] & ratio to NLO \\ 
\midrule
NLO+PS ($m_{b\bar{b} \ell\ell}$)\;& \; $31.86(1)_{-13.3\%}^{+16.3\%}$\,& 1.000 \\
\minlo{} ($m_{b\bar{b} \ell\ell}$)\;& \; $22.33(1)_{-17.9\%}^{+28.2\%}$\,& 0.701 \\
\minnlo{} ($m_{b\bar{b} \ell\ell}$) \;& \; $50.58(4)_{-12.2\%}^{+16.8\%}$\, & 1.588 \\ 
\midrule
NLO+PS ($H_T/2$)  \;& \; $41.42(1)_{-15.4\%}^{+19.2\%}$\,& 1.000\\
\minnlo{} ($H_T/2$)  \;& \;  $57.68(5)_{-12.9\%}^{+18.3\%}$\,& 1.393\\ 
\bottomrule
\end{tabular}
\end{center}
\vspace{-1em}
  \caption{
    Total $b \bar b Z$ cross section with $66\,{\rm GeV}\le m_{\ell^+\ell^-}\le 116\,{\rm GeV}$. The scale in brackets indicates the different 
   scale setting as described in the text.
    The quoted errors represent 
    scale uncertainties, while the numbers in brackets are  
    numerical uncertainties on the last digit.\label{tab:XS}}
\end{table}

Table\,\ref{tab:XS} shows the $pp\to b\bar{b}\ell^+\ell^-$ total cross section. 
For reference, NLO+PS (\minnlo{}) results with a central scale $H_T/2$ ($\mu_R^{(0)}= H_T/2$)
are given as well, where $H_T$ is the sum over the transverse masses of each bottom quark and each lepton.
Shower effects are negligible for the inclusive rate and we keep
effects from hadronization, multi-parton interactions (MPI) and QED radiation off.

The \minlo{} prediction, which is formally NLO accurate and includes
additional $\mathcal{O}(\alpha_s^2)$ corrections, is 43\% smaller than the other NLO+PS result and 
does not provide a reasonable prediction. 
Although this may seem surprising, this behaviour can be explained by 
the substantial cancellation of logarithmic corrections in $m_b$ between
the real (double-real and real-virtual) and the double-virtual amplitudes.
Those logarithmic terms originate from the massive bottom quark in the 4FS, which 
regulates the real phase-space integration as well as 
 the loop integration. The ensuing logarithmic contributions are bound 
to cancel between real and double-virtual amplitudes, which can be understood by
considering the 5FS, where these logarithms would appear as
$1/\epsilon$ poles that cancel by the KLN theorem~\cite{Kinoshita:1962ur,Lee:1964is}.

For \minlo{}, the relative $\mathcal{O}(\alpha_s^2)$ contribution is incomplete as 
only the real amplitudes are included and the corresponding logarithmic terms 
induce a numerically significant negative effect. 
We have checked that it is sufficient to include the logarithmic 
corrections in the double-virtual amplitudes 
(obtained using the massification procedure in \app{sec:finite_massification})
to restore the appropriate cancellation and obtain a positive $\mathcal{O}(\alpha_s^2)$ correction.
Due to this unphysical effect we refrain from including \minlo{} results in the remainder of this letter.

\begin{table}[b!]
  \vspace*{0.3ex}
  \begin{center}
\begin{tabular}{l|ll}
\toprule
$\sigma_{\rm fiducial}$ [pb] & \multicolumn{1}{c}{$Z$+$\ge 1$ $b$-jet} & \multicolumn{1}{c}{$Z$+$\ge 2$ $b$-jets} \\
\midrule
NLO+PS (5FS)~\cite{CMS:2021pcj} & $\phantom{00}7.03 \pm 0.47\phantom{0}$          & $\phantom{00}0.77 \pm 0.07\phantom{0}$ \\
NLO+PS (4FS)                         & $\phantom{00}4.08 \pm 0.66\phantom{0}$          & $\phantom{00}0.44 \pm 0.08\phantom{0}$ \\
\minnlo{}  (4FS)             & $\phantom{00}6.72 \pm 0.91\phantom{0}$          & $\phantom{00}0.79 \pm 0.10\phantom{0}$ \\
CMS~\cite{CMS:2021pcj}          & $\phantom{00}6.52 \pm 0.43\phantom{0}$          & $\phantom{00}0.65 \pm 0.08\phantom{0}$ \\
\bottomrule
\end{tabular}
\end{center}
\vspace{-1em}
  \caption{Comparison of theory predictions with the fiducial cross-section 
  measurements by CMS~\cite{CMS:2021pcj} for (at least) one and two tagged $b$-jets. 
  Experimental uncertainties are added in quadrature. The 5FS result
  is taken from \citere{CMS:2021pcj} and  scale uncertainties 
  have been symmetrized by taking the 
  maximum of the absolute value of the errors.}
\label{tab:data}
\end{table}

Considering the \minnlo{} predictions in \tab{tab:XS}, NNLO corrections 
increase the NLO cross section by 59\%, which renders
them crucial for an accurate prediction in the 4FS.
Since the NNLO corrections are much larger than the NLO scale uncertainties,
we consider $H_T/2$ as a second scale choice. In this case,  
the NLO and the NNLO cross sections are larger, with NNLO corrections of about 39\%. 
It is reassuring that the dependence on the scale choice reduces at NNLO compared to NLO.

In Table\,\ref{tab:data} we consider the fiducial cross section measurement of the most recent 
CMS analysis for $Z$-boson production in association with bottom-flavoured jets ($b$-jets)
\cite{CMS:2021pcj}, which includes the complete Run-2 data set of the LHC. 
The fiducial cuts are defined in \app{app:cuts}.
To warrant a realistic comparison to data we include
effects from hadronization, multi-parton interactions (MPI) and QED radiation from now on.
In our 4FS calculation, we can directly apply the experimental definition of \mbox{$b$-jets}, which 
is rendered infrared safe by the bottom mass, while in the 5FS the
naive definition of $b$-jets leads to divergences \cite{Banfi:2006hf,Caletti:2022hnc,Czakon:2022wam,Gauld:2022lem,Caola:2023wpj}.
For the $Z$+$\ge$1(2) $b$-jet rate we observe a clear tension with the NLO+PS result in the 4FS, 
whose central cross section is $40(20)\%$ below the measurement, well outside the uncertainties. 
By contrast, the \minnlo{} prediction is in full agreement with the data within uncertainties.

In addition, we include in Table\,\ref{tab:data} the NLO+PS prediction in the 5FS, as quoted in the CMS analysis~\cite{CMS:2021pcj} and obtained with {\sc MadGraph5\_aMC@NLO}~\cite{Alwall:2014hca}.\footnote{We note that several NLO+PS predictions are quoted in \citere{CMS:2021pcj}. The one quoted in Table\,\ref{tab:data} is the only one using the same PDF set as in our setup.}
NLO+PS results in 4FS and 5FS are not compatible with each other, but
our \minnlo{} prediction agrees with the 5FS result, thanks to  the inclusion of NNLO QCD corrections in the 4FS.
While both predictions agree similarly well with the $Z$+$\ge$2 $b$-jets data, \minnlo{}
shows a better description in the $Z$+$\ge$1 $b$-jet fiducial region.

 \begin{figure*}[t]
     \centering
     \begin{tabular}{ccccc}
   \includegraphics[width=0.32\textwidth,page=1]{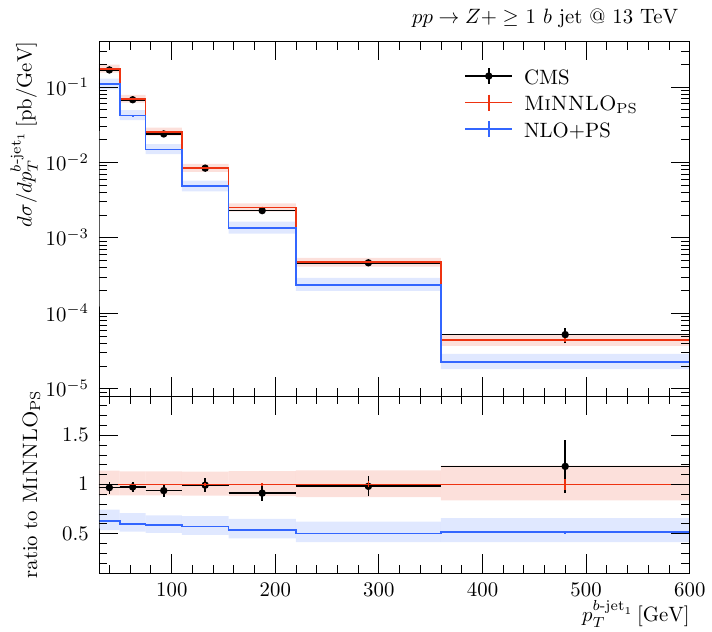} &&%
   \includegraphics[width=0.32\textwidth,page=2]{plots_paper.pdf} &&%
   \includegraphics[width=0.32\textwidth,page=3]{plots_paper.pdf}\\%
   \includegraphics[width=0.32\textwidth,page=4]{plots_paper.pdf} &&%
   \includegraphics[width=0.32\textwidth,page=5]{plots_paper.pdf} &&%
   \includegraphics[width=0.32\textwidth,page=6]{plots_paper.pdf}\\%
      \end{tabular}
      \caption{ \label{fig:dist} Comparison of theory predictions with differential distributions measured 
      by CMS~\cite{CMS:2021pcj}.}
 \end{figure*}

Next, we examine a set of differential distributions in Fig.\,\ref{fig:dist} and compare
NLO+PS (blue, dashed) and \minnlo{} (red, solid) predictions to the CMS measurement (black data points) \cite{CMS:2021pcj}. 
The first three figures show observables in the inclusive 1-$b$-jet phase space: 
the transverse momentum ($p_T^{b\text{-jet}_1}$) and pseudo-rapidity ($\eta^{b\text{-jet}_1}$) of the leading $b$-jet as well as 
the distance between the $Z$ boson and the leading $b$-jet in the $\eta$-$\phi$-plane ($\Delta R^{Z, b\text{-jet}_1}$).
The other three figures show distributions in the  inclusive 2-$b$-jet phase space:
the transverse momentum of the leading ($p_T^{b\text{-jet}_1}$) and 
subleading $b$-jet ($p_T^{b\text{-jet}_2}$), and the invariant mass of the two $b$-jets ($m_{bb}$).

As observed for the fiducial rates, NLO+PS predictions fail in describing 
the normalization of the measured cross sections. Additionally, the shapes of 
some of the distributions are not reproduced particularly well.
The \minnlo{} predictions, on the other hand, are in agreement with data, especially
for  Z+$\ge$1-$b$-jet observables the central prediction is 
almost spot on the data points, which is remarkable 
considering the relatively high precision of theory predictions and measurement.
In the inclusive 2-$b$-jet phase space the experimental errors are larger due to 
the lower statistics. Here, \minnlo{} predicts a normalization slightly higher than the data 
points, but still fully covered by the uncertainties, while the differential shapes are reproduced  
particularly well.

An exception to the previous statements about \minnlo{} is the difference observed at high $\Delta R^{Z, b\text{-jet}_1}$ 
with the CMS data, which originates from large values of the rapidity separation between the $Z$ boson
and the leading $b$-jet ($\Delta y^{Z, b\text{-jet}_1}$), where similar differences appear. 
Although less pronounced than in the 4FS, such trend is also present the 5FS predictions at large 
$\Delta y^{Z, b\text{-jet}_1}$ and $\Delta R^{Z, b\text{-jet}_1}$, see Figs.\,6 and 7 of \citere{CMS:2021pcj}.
That behaviour also appears in the $\Delta y^{Z, b\text{-jet}_1}$ distribution in an earlier ATLAS measurement \cite{ATLAS:2020juj}. 
A better understanding of this discrepancy requires additional studies and potentially all-order resummation 
of logarithms in $m_b$ through a 4FS and 5FS combination at NNLO+PS, which is left for future work.

\vspace{0.5cm}{\it Summary.---}%
In this letter, we presented a novel computation for the production of a $Z$-boson in association
with bottom quarks in hadronic collisions. We have calculated NNLO QCD corrections in the 4FS, 
including the five-point two-loop amplitude in the small-$m_b$ approximation.
In addition, the first NNLO+PS approach for the production of a heavy-quark pair in association with colour-singlet particles has been developed,
which can be readily applied to other processes, like $b\bar{b} W$~\cite{Buonocore:2022pqq},  $t\bar{t} W$~\cite{Buonocore:2023ljm}, and  $t\bar{t} H$~\cite{Catani:2022mfv} production. 

Our NNLO+PS calculation solves two (related) long-standing issues for $b\bar{b}Z$ production:
First, the significant tension of NLO(+PS) predictions in the 4FS with experimental data. 
Second, the large differences between 4FS and 5FS predictions for this process~\cite{Krauss:2016orf}.
Our analysis identifies that missing higher-order corrections in the 4FS cause these discrepancies
and that the perturbative accuracy of previous calculations has been insufficient. 
Including NNLO QCD corrections brings the 4FS predictions in agreement with the experimental data and with the 5FS results. 
The calculation presented in this letter also builds the basis for a more accurate determination of the bottom-quark
mass effects in Drell-Yan production, relevant for $M_W$ measurements, along the lines of the study in \citere{Bagnaschi:2018dnh},
which at the time was pursued only at NLO+PS.
\\

{\it Acknowledgements.---}%
 We would like to thank 
 Luca Buonocore, Fernando Febres Cordero, Rhorry Gauld,
 Massimiliano Grazzini, Pier Francesco Monni, 
 Luca Rottoli and Giulia Zanderighi, 
 for fruitful discussions and comments on the manuscript.  
 We are indebted to
 Federico Buccioni for providing {\sc OpenLoops} amplitudes with a different number of quarks running in the loops.
 We are thankful to Stefan Kallweit for helping us with a fixed-order implementation, which we used for comparison.
 We further thank Luca Buonocore and Luca Rottoli for performing cross checks on the massification procedure with us,
 and we thank Chiara Savoini and Massimilano Grazzini for bringing to our attention the issues related to the singular behaviour of the 
 massless amplitudes for certain 
 mappings from the massive to the massless momenta of the bottom quarks. 

\bibliography{MiNNLO_bbZ.bib}

\input{supplementary_material}

\end{document}

%% file: feynman_diagrams.tex
 \resizebox{0.48\textwidth}{!}{
\begin{tikzpicture}
\begin{feynman}
	\vertex (a1) at (0,0) {\( q\)};
	\vertex (a2) at (0,-2) {\(\bar q\)};
	\vertex (a3) at (1.0,-1);
	\vertex (a4) at (2.0,-1);
	\vertex (a40) at (2.3,-1.3);
	\vertex (a41) at (2.7,-1.1);
	\vertex (a42) at (3.5,-0.6){\( \ell^+\)};
	\vertex (a43) at (3.5,-1.6){\( \ell^-\)};
	\vertex (a5) at (3.0,0){\( b\)};
	\vertex (a6) at (3.0,-2){\(\bar b\)};
        \diagram* {
          {[edges=fermion]
            (a1)--[thick](a3)--[thick](a2),
            (a6)--[fermion, ultra thick](a40)--[fermion, ultra thick](a4)--[fermion, ultra thick](a5),
            (a42)--[thick](a41)--[thick](a43)
          },
          (a3) -- [gluon,thick] (a4),
          (a40)--[boson,thick](a41),
        };
      \end{feynman}
\end{tikzpicture}
\begin{tikzpicture}
  \begin{feynman}
	\vertex (a1) at (0,0) {\( g\)};
	\vertex (a2) at (0,-2) {\( g\)};
	\vertex (a3) at (1.0,-1);
	\vertex (a4) at (2.0,-1);
	\vertex (a40) at (2.3,-1.3);
	\vertex (a41) at (2.7,-1.1);
	\vertex (a42) at (3.5,-0.6){\( \ell^+\)};
	\vertex (a43) at (3.5,-1.6){\( \ell^-\)};
	\vertex (a5) at (3,0){\( b\)};
	\vertex (a6) at (3,-2){\(\bar b\)};
        \diagram* {
          {[edges=fermion]
            (a6)--[fermion, ultra thick](a40)--[fermion, ultra thick](a4)--[fermion, ultra thick](a5),
            (a42)--[thick](a41)--[thick](a43)
          },
          (a3) -- [gluon,thick] (a4),
          (a2)--[gluon,thick](a3)--[gluon,thick](a1),
          (a40)--[boson,thick](a41),
        };
  \end{feynman}
\end{tikzpicture}
\begin{tikzpicture}
  \begin{feynman}
	\vertex (a1) at (0,0) {\( g\)};
	\vertex (a2) at (0,-1.7) {\( g\)};
	\vertex (a3) at (1.3,0);
	\vertex (a32) at (1.3,-0.85);
	\vertex (a33) at (2.,-0.85);
	\vertex (a34) at (2.9,-0.4){\( \ell^+\)};
	\vertex (a35) at (2.9,-1.3){\( \ell^-\)};
	\vertex (a4) at (1.3,-1.7);
	\vertex (a5) at (2.6,0){\( b\)};
	\vertex (a6) at (2.6,-1.7){\(\bar b\)};
        \diagram* {
          {[edges=fermion]
            (a6)--[fermion, ultra thick](a4)--[fermion, ultra thick](a32)--[fermion, ultra thick](a3)--[fermion, ultra thick](a5),
            (a34)--[thick](a33)--[thick](a35)
          },
          (a2)--[gluon,thick](a4),
          (a3)--[gluon,thick](a1),
          (a32)--[boson,thick](a33),
        };
  \end{feynman}
\end{tikzpicture}}
\caption{\label{fig:diagrams} Feynman diagrams for 
  $pp\to b\bar{b}\ell^+\ell^-$ at LO.}

%% file: supplementary_material.tex
\onecolumngrid
\appendix

\section*{Supplemental material}
\newlength{\fwidth}
\setlength{\fwidth}{0.3\textwidth}
   
\section{Finite massification}
\label{sec:finite_massification}

\newcommand{\lmb}{~\ell_b}
\newcommand{\CF}{C_F}
\newcommand{\CA}{C_A}
\newcommand{\Nh}{n_h}
\newcommand{\Nf}{n_f}
\newcommand{\Nl}{n_l}

In this appendix we provide additional details on the small bottom mass approximation (exploiting a massification procedure) that we use to compute the two-loop amplitude for $b\bar{b}Z$ production.
The massification procedure allows us to approximate a massive amplitude from the massless one, while dropping power corrections in the bottom mass $m_b$, but 
reproducing correctly all logarithmic terms in $m_b$.

We aim to calculate the finite remainder $\ket{\mathcal{R}} = \mathbf{Z}^{-1} \ket{\mathcal{M}}$ defined in \eqn{eq:hard_function}
(we leave the channel labels implicit for clarity of the formulae) in the small bottom mass approximation.
Using the factorization formula from \citeres{Mitov:2006xs,Wang:2023qbf}, we obtain
\begin{equation} \label{eq:finite_massification}
  \ket{\mathcal{R}_{m_b\ll \mu_h}} = \mathbf{Z}^{-1}_{m_b\ll \mu_h}\,\mathcal{F}\,\mathbf{S} \, \eval{\mathbf{Z}_0 \, \ket{\mathcal{R}_{0}}}_{\Nf = \Nl + \Nh}\,, \quad \text{with}\quad \ket{\mathcal{R}} = \ket{\mathcal{R}_{m_b\ll \mu_h}}  + \order{\frac{m_b}{\mu_h}}\, ,
\end{equation}
where $\mu_h$ is the characteristic hard scale of the process, 
$\ket{\mathcal{R}_{0}} = \mathbf{Z}^{-1}_{0} \ket{\mathcal{M}_{0}}$ is the massless finite remainder defined through the \emph{massless} IR renormalization operator $\mathbf{Z}_0$ \cite{Becher:2009cu,Becher:2009qa},
and $\mathbf{Z}_{m_b\ll \mu_h}$ is the small mass expansion of $\mathbf{Z}$. The channel-dependent operators $\mathcal{F}$ are defined as
\begin{equation}
   \mathcal{F}_{q\bar{q}} = \mathcal{Z}_{[Q]} \mathcal{Z}_{[q]}, \qquad \mathcal{F}_{gg} = \mathcal{Z}_{[g]} \mathcal{Z}_{[q]}\, ,
\end{equation}
with the functions $\mathcal{Z}_{[i]}$ given in \citeres{Mitov:2006xs,Wang:2023qbf}, and $\mathbf{S}$ is the soft function introduced in \citere{Wang:2023qbf}
that is required to approximate contributions from heavy quark loops. 
Here $\mathbf{Z}_0$ and $\ket{\mathcal{R}_{0}}$ depend on the total number of flavors $\Nf=\Nl+\Nh$, which is determined by the number of $\Nl=4$ light flavours in the 4FS 
and $\Nh=1$ heavy flavours corresponding to the bottom quark. Since the top quark practically decouples due to its large mass, its loop corrections are expected to be 
negligible. Therefore, we discard contributions from top loops here.

We  introduce functions $\bar{\mathcal{F}}$ and $\bar{\mathbf{S}}$ free from $\epsilon$ poles, 
such that $\bar{\mathcal{F}}\, \bar{\mathbf{S}} = \mathbf{Z}^{-1}_{m_b\ll \mu_h}~\mathcal{F}~\mathbf{S}~\mathbf{Z}_0$.
Here, $\bar{\mathcal{F}}$ is diagonal in colour space, while $\bar{\mathbf{S}}$ is a matrix in colour space.
We start by writing their perturbative expansions in the strong coupling constant $\as^{\Nf}$, which includes the heavy flavours in its running,
\begin{align} \label{eq:massification_coeffs_start}
  \bar{\mathcal{F}} &= 1 + \left(\frac{\as^{\Nf}}{2\pi}\right) \bar{\mathcal{F}}^{(1)} + \left(\frac{\as^{\Nf}}{2\pi}\right)^2 \bar{\mathcal{F}}^{(2)} + \mathcal{O}(\as^3)\, ,  \\
  \bar{\mathbf{S}} &= 1 + \qty(\frac{\as^{\Nf}}{2\pi})^2 \mathbf{C}_d ~ \mathcal{S}^{(2)} + \mathcal{O}(\as^3), \qquad \mathbf{C}_d = \sum_{(i,j)} -\frac{\mathbf{T_i} \cdot \mathbf{T_j}}{2} \log\left(\frac{-s_{ij}}{\mu_R^2}\right)\, ,
\end{align}
where $\mathbf{C}_d$ is the standard dipole colour-space operator (see e.g.\ \citere{Becher:2009qa}), and the coefficients are 
\begin{align}
  \bar{\mathcal{F}}^{(1)} &= 2 \CF \lmb^{2} + \CF \lmb + \CF \qty( 2 + \frac{\pi^{2}}{12}  ) ~+~ \Nh \bar{\mathcal{F}}^{(1)}_{\Nh,c\bar c}\, ,\quad \text{with}\quad  \bar{\mathcal{F}}^{(1)}_{\Nh,q\bar{q}} = 0\, , \quad \bar{\mathcal{F}}^{(1)}_{\Nh,gg}= -\frac{2}{3} \lmb\, ,\\
  \nonumber
  \bar{\mathcal{F}}^{(2)} &=   2 \CF^{2} \lmb^{4} + \lmb^{3} \qty( \frac{22}{9} \CA \CF + 2 \CF^{2} - \frac{4}{9} \CF \Nl ) + \\ \nonumber
  \lmb^{2} &\qty(\CA \CF \qty( \frac{167}{18} - \frac{\pi^{2}}{3}) + \CF^{2} \qty( \frac{9}{2} + \frac{\pi^{2}}{6}  ) -\frac{13}{9} \CF \Nl ) + \\ \nonumber 
  \lmb &\qty( \CA \CF \qty( \frac{1165}{108} + \frac{14}{9} \pi^{2} - 15 \zeta_3 ) + \CF^{2} \qty( \frac{11}{4} - \frac{11}{12} \pi^{2} + 12 \zeta_3 ) - \CF \Nl \qty( \frac{77}{54} + \frac{2}{9} \pi^{2} )) +   \\ \nonumber
  \phantom{\lmb} &  \CF^{2} \qty( \frac{241}{32} - \frac{163}{1440} \pi^{4} + \pi^{2} \qty( \frac{13}{12} - 2 \log (2) ) - \frac{3}{2} \zeta_3 ) + \CA \CF \qty( \frac{12877}{2592} - \frac{47}{720} \pi^{4} + \pi^{2} \qty( \frac{323}{432} + \log (2) ) + \frac{89}{36} \zeta_3 ) -  \\ 
  \phantom{\lmb} & \CF\, \Nl \qty( \frac{1541}{1296} + \frac{37}{216} \pi^{2} + \frac{13}{18} \zeta_3 ) \;+\; \Nh\, \bar{\mathcal{F}}^{(2)}_{\Nh,c \bar{c}}, 
\end{align}
with
\begin{align}
  &\bar{\mathcal{F}}^{(2)}_{\Nh,q\bar{q}} = - \frac{20}{9} \lmb^{3} \CF  + \frac{32}{9} \lmb^{2} \CF  + \lmb \CF \qty( -\frac{157}{27} - \frac{7}{18} \pi^{2} ) + \CF \qty( \frac{1933}{324} - \frac{13}{108} \pi^{2} - \frac{7}{9} \zeta_3 ),  \\ \nonumber
  &\bar{\mathcal{F}}^{(2)}_{\Nh,gg} =  \lmb^{3} \qty( -\frac{4}{9} \CA - \frac{28}{9} \CF ) + \lmb^{2} \qty( \frac{10}{9} \CA + \frac{7}{9} \CF ) + \lmb \qty( \CF \qty( -\frac{319}{54} - \frac{5}{18} \pi^{2} ) + \CA \qty( -\frac{92}{27} + \frac{\pi^{2}}{18}  ) )  +  \\ 
  &\phantom{\bar{\mathcal{F}}^{(2)}_{\Nh,gg}} \CA \qty( \frac{179}{108} - \frac{5}{216} \pi^{2} - \frac{7}{18} \zeta_3 ) + \CF \qty( \frac{2677}{1296} - \frac{41}{216} \pi^{2} - \frac{\zeta_3}{18}  )  ~+~ \frac{4}{9} \Nh \lmb^2\,,
\end{align}
  and
\begin{align}\label{eq:massification_coeffs_end}
  \mathcal{S}^{(2)} &=  \Nh \qty( - \frac{2}{3} \lmb^{2} + \frac{10}{9} \lmb -\frac{14}{27} )\, ,
\end{align}
where $\lmb = -\log\left( m_b/\mu_R \right)$.
To match the implementation of the amplitudes in a 4FS calculation,
we then convert the expansion in $\as^{\Nf}$ of $\bar{\mathcal{F}}$, $\bar{\mathbf{S}}$, and the squared finite remainders into an expansion in the coupling constant $\as^{\Nl}$ of the effective theory with $\Nh$ decoupled quarks \cite{Bernreuther:1981sg} through  a finite renormalization shift
\begin{align}\label{eq:as_decoupling}
  \as^{\Nf} &= \as^{\Nl}\left\{ 1 + \qty(\frac{\as^{\Nl}}{2\pi}) \frac{2}{3} \lmb \Nh + \qty(\frac{\as^{\Nl}}{2\pi})^{2} \left[  \frac{4}{9} \lmb^2 \Nh^2 + \lmb \Nh \qty( \frac{5}{3} \CA + \CF ) + \Nh \qty( -\frac{4}{9} \CA + \frac{15}{8} \CF )   \right] \right\} \,.
\end{align}
We note that in \citere{Ferroglia:2009ii} $\mathbf{Z}^{-1}$ is defined in the effective theory with $\Nh$ decoupled quarks.
To obtain the expansions in $\as^{\Nf}$ in Eqs.~\eqref{eq:massification_coeffs_start} to \eqref{eq:massification_coeffs_end}
, we expand $\mathbf{Z}^{-1}$ in $\as^{\Nf}$ by applying the inverse of the decoupling relation \eqref{eq:as_decoupling} \cite[Section 4.2]{Ferroglia:2009ii}.

Finally, we calculate the two-loop contribution as
  \begin{align}\label{eq:twoloop}
    2 \Re\braket{\mathcal{R}^{(0)}}{\mathcal{R}^{(2)}} &= \frac{\abs{\mathcal{R}^{(0)}}^2}{\abs{\mathcal{R}^{(0)}_{0}}^2} ~ 2 \Re\braket{\mathcal{R}^{(0)}_0}{\mathcal{R}^{(2)}_{m_b\ll \mu_h}} + \order{\frac{m_b}{\mu_h}}\, , \\
    \label{eq:M2M0}
    2 \Re\braket{\mathcal{R}^{(0)}_0}{\mathcal{R}^{(2)}_{m_b\ll \mu_h}} & =\\ \nonumber
    &2 \bar{\mathcal{F}}^{(2)} \abs{\mathcal{R}^{(0)}_{0}}^2  + \bar{\mathcal{F}}^{(1)} 2 \Re\braket{\mathcal{R}^{(0)}_0}{\mathcal{R}^{(1)}_{0}} + \mathcal{S}^{(2)} 2\Re\ev{\mathbf{C}_d}{\mathcal{R}^{(0)}_0} +
    2 \Re\braket{\mathcal{R}^{(0)}_0}{\mathcal{R}^{(2)}_{0}} \, .
  \end{align}
Here $X^{(i)}$ with $X=\{\mathcal{R,\bar{\mathcal{F}},\mathcal{S}}\}$ are now the coefficients in the expansion in $\as^{\Nl}$,
derived from their expressions expanded in $\as^{\Nf}$ above and the expansion of the squared finite remainders in $\as^{\Nf}$
using the decoupling relation in \eqn{eq:as_decoupling}.
The logarithmic contributions in $m_b$ are fully contained within the first three terms of \eqn{eq:M2M0}, while the two-loop diagrams are contained in the massless finite remainder at second order
$\Re\braket{\mathcal{R}^{(0)}_0}{\mathcal{R}^{(2)}_{0}}$, which we obtain from the analytic results of \citere{Abreu:2021asb}.
Since the results of \citere{Abreu:2021asb} are obtained in the ``Catani scheme" \cite{Catani:1998bh}, 
we perform a finite renormalization shift (see e.g.\ \citere{DeLaurentis:2023izi}) to convert them into the scheme employed here.
The coefficients $\kappa_{c \bar{c}, i}$ of $\log\left( m_b/\mu_R \right)$ powers in \eqn{eq:mbexpansion} are then obtained 
from the first three terms on the right hand side of \eqn{eq:M2M0} using the results of Eqs.~\eqref{eq:massification_coeffs_start} to \eqref{eq:as_decoupling}.

\section{Momentum mappings}
\label{sec:momentum_maps}

In our 4FS calculation, the phase-space integration is performed assuming massive bottom quarks with $m_b\neq 0$. However,
the massless finite remainder $\mathcal{R}_{0}$ entering \eqn{eq:twoloop} 
must be evaluated on on-shell phase-space points $P_0$ with $m_b=0$.
We therefore need an explicit mapping of massive phase-space points $P$, $\eta : P \to  P_0$, such that  $\eta(P) = P_0 + \order{m_b/\mu_h}$.
In addition, we have to ensure that $\eta$ does not cause $\mathcal{R}_{0}$ to be evaluated near its singularities.
Since the quark- and gluon-initiated channels have distinct leading order momentum flows (see \fig{fig:diagrams}),
it is useful to employ dedicated mappings $\eta_{q \bar{q}}, \eta_{gg}$ for each of the channels.

Let us first define our notation of the momenta by introducing the process as
\begin{align}
p(p_1)\,p(p_2)\to  b(p_b)\,\bar{b}(p_{\bar b})\, \ell^+(p_{\ell^+})\,\ell^-(p_{\ell^-})\,.
\end{align}
For $\eta_{q \bar{q}}$ we perform the simultaneous light-cone decomposition \cite{vanHameren:2005ed} of the massive bottom and anti-bottom momenta $p_b$ and $p_{\bar{b}}$,
respectively, and determine the massless momenta $\hat p_b$ and $\hat p_{\bar{b}}$ as
\begin{align} \nonumber
  \hat{p}_{b} &= \alpha^+ p_b - \alpha^- p_{\bar{b}}, \qquad \alpha^{\pm} = \frac{1}{2} \qty(1 \pm \qty(1 - 4 \frac{m_b^2}{m_{b \bar{b}}})^{-\frac{1}{2}}),  \\
  \hat{p}_{\bar{b}} &= \alpha^+ p_{\bar{b}} - \alpha^- p_b,
\end{align}
which preserves the total momentum $\hat p_{b \bar{b}}\equiv p_{b \bar{b}}$ of the $b \bar{b}$ system and prevents a collinear $g\to b\bar{b}$ splitting in 
the quark channel. 
The mapping $\eta_{q \bar{q}}$ is minimal in the sense that only the bottom-quark momenta are modified.

An undesirable side effect of the mapping $\eta_{q \bar{q}}$ (when applied in the gluon channel) 
is that $\hat{p}_{b}$ or $\hat{p}_{\bar{b}}$ can become collinear to the initial state momenta $p_1$ or $p_2$
when the $b \bar{b}$ pair is produced at the threshold.
In the gluon channel this introduces a collinear singularity, and we therefore construct $\eta_{gg}$ such that it avoids these configurations.
First, we set the massless momenta to
\begin{align}
  \hat{p}_{x} &= p_{x} + \qty(\sqrt{ 1 - \frac{m_b^2 n_x^2}{(p_x \cdot n_x)^2}} - 1) \frac{(p_x \cdot n_x)}{n_x^2} ~ n_x \quad\text{with }  x \in \{b, \bar{b}\},\\
  \label{eq:mapgg}
   n_{x} &= p_{x} - p_1 \frac{(p_2 \cdot p_x)}{(p_1 \cdot p_2)}- p_2 \frac{(p_1 \cdot p_x)}{(p_1 \cdot p_2)}, 
\end{align}
where $n_{x}$ are transverse to both $p_1$ and $p_2$. 
Then to restore momentum conservation we consider two options: Either
we redistribute $\Delta p_{b \bar{b}} = p_b + p_{\bar{b}} - \hat{p}_b - \hat{p}_{\bar{b}}$
into $\hat{p}_1$ and $\hat{p}_2$, such that $\hat{p}_{12} = \hat{p}_1+\hat{p}_2 = p_1 + p_2 - \Delta p_{b \bar{b}}$, by 
performing a Lorentz boost on $p_1$ and $p_2$ in the direction $-\hat{p}_{12}$ followed by rescaling with $\sqrt{\hat{p}_{12}^2/p_{12}^2}$. 
Or, we redistribute $\Delta p_{b \bar{b}}$ into the lepton momenta  $\hat{p}_{\ell^+}$ and $\hat{p}_{\ell^-}$ instead.
We have verified that in both cases $\eta_{gg}$ avoids the collinear singularities in the massless amplitudes, but we choose the first
option as our default choice, since it leaves the $Z$ momentum untouched.
Comparing the two choices, we estimated the uncertainty introduced by the mapping in the gluon channel to be at the sub-percent level
and roughly of the order of the numerical error of our calculation.

\section{Fiducial phase-space definition}
\label{app:cuts}

In this Appendix we specify the set of fiducial cuts defined in the CMS analysis of \citere{CMS:2021pcj} and used in this letter for the
comparison to data.
The final state under consideration containes a $Z$ boson, decaying either to electrons or muons, and at least one (or two) $b$-jets.

The $Z$ boson candidate is reconstructed from two same-flavour leptons with opposite charge, with their invariant mass $m_{\ell^+\ell^-}$ in the window $71\,{\rm GeV}\le m_{\ell^+\ell^-}\le 111\,{\rm GeV}$. The leptons are required to satisfy the constraint on their transverse momentum of $p_T^{\ell}>25$~GeV, and a requirement on their pseudo-rapidity of $|\eta^\ell|<2.4$. Events with more than two leptons satisfying these cuts are vetoed. In addition, the leading lepton must satisfy the constraint $p_T^{\ell_1}>35$~GeV. The leptons are ``dressed'' by adding the momenta of all photons within a radius of $\Delta R^{\gamma,\ell} \leq 0.1$.

Jets are defined by clustering all light partons plus the bottom quark using the anti-$k_T$ algorithm with a radius of $R=0.4$. They
are classified as $b$-jets if they contain at least one bottom-flavoured hadron and if they fulfill the transverse-momentum and pseudo-rapidity thresholds of $p_T^{b\text{-jet}}>30$\,GeV and $|\eta^{b\text{-jet}}|<2.4$, respectively.
In the $Z$+$\ge 1$ $b$-jet and $Z$+$\ge 2$ $b$-jets fiducial regions a reconstructed $Z$ boson and at least one or two $b$-jets are required,
respectively. The overlap between the leptons (from the $Z$ boson decay) and the $b$-jets 
is removed by requiring a minimum distance of $\Delta R^{\ell,b\text{-jet}}>0.4$ between them.